\documentstyle[manuscript,osa]{revtex}

\begin{document}
\title{EFFECTS OF\ THREE-BODY INTERACTIONS\ ON\ THE\ STRUCTURE\ AND THERMODYNAMICS
OF LIQUID KRYPTON }
\author{N. Jakse, J.M. Bomont and J.L. Bretonnet}
\address{Laboratoire de Th\'{e}orie de la Mati\`{e}re Condens\'{e}e, \\
Universit\'{e} de Metz, 1, bd F. D. Arago, CP 87811, \\
57078 Metz Cedex 3, France}
\maketitle

\begin{abstract}
Large-scale molecular dynamics simulations are performed to predict the
structural and thermodynamic properties of liquid krypton using a potential
energy function based on the two-body potential of Aziz and Slaman plus the
triple-dipole Axilrod-Teller (AT) potential. By varying the strength of the
AT potential we study the influence of three-body contribution beyond the
triple-dipole dispersion. It is seen that the AT potential gives an overall
good description of liquid Kr, though other contributions such as higher
order three-body dispersion and exchange terms cannot be ignored.
\end{abstract}

\baselineskip=20pt

The knowledge of interactions in noble gases remains a fundamental question
that is not completely solved. Despite the simplicity of their closed-shell
electronic structure, it is well-known that a simple pair potential, though
giving the essential features of the structural and thermodynamic
properties, is not sufficient for a quantitative description, and many-body
effects have to be taken into account \cite{Bar1976}. Significant advances
have been made when it has been demonstrated \cite{End1965,Rea1992} that the
static structure factor $S(k)$ at small wave-number, $k$, is directly
related to the long range part of the effective potential between pairs of
atoms. It was therefore recognized that precise measurements of $S(k)$ could
provide a direct observation of the details of the interactions. During the
past few years, high precision experiments \cite
{Mag1996,For1997,For1998,Ben1999,Gua2001}, performed in the range $0.5<k<4$
nm$^{-1}$ using small angle neutron scattering facilities, have confirmed
the presence of an additional interaction at large distance that can be
associated to a three-body contribution at least for Kr \cite{For1997} and
Xe \cite{For1998}.

Long range interactions in noble gases arising from induced dipoles are
known as the London dispersion forces \cite{Lon1930}, and are expected to
behave as $r^{-6}$, where $r$ is the interatomic distance. Quantum
electromagnetic effects acting at very large separations \cite{Pow2001},
called retardation effects, are described by the Casimir-Polder potential 
\cite{Cas1948}, which is expected to fall off as $r^{-7}$ and to have a
negligibly small influence \cite{Rea1992} on $S(k)$. It is not surprising
that a simple Lennard-Jones (12-6) potential is a satisfactory effective
pair potential at first sight since it can be seen as including mean effects
coming from the different dipole-dipole, multipole-multipole as well as
higher order terms. However, a better understanding of the interactions lies
in a careful examination of the various genuine contributions of the
multipole expansion, which contains two-body $dd$, $dq$, $qq$ terms, etc.,
as well as irreducible three-body contributions such as $ddd$, $ddq$, $dqq$, 
$qqq$, etc., where $d$ and $q$ denote the dipole and quadrupole moment,
respectively. According to Barker and Henderson \cite{Bar1976}, four --and
more--\ body terms are very small and can be neglected. The two-body
potential of noble gases is often taken in the very accurate
Hartree-Fock-dispersion form propounded by Aziz and Slaman \cite
{Azi1985,Azi1986}, while three-body dispersion effects are often represented
by the $ddd$ term in the form given by Axirod and Teller \cite{Axi1943} that
represents the major contribution. The other three-body contributions
mentioned above, which can be modelled by the expressions of Bell \cite
{Bel1970}, are known to have a small influence although it is not clear
whether the agreement with the experiment could be improved when these are
taken into account. At high densities, beside dispersion terms, three-body
overlap contributions such as exchange effects may come into play \cite
{Cop1968,Lou1988,Sad1996,Lot1997}.

It is tempting to consider an effective three-body potential of the
Axirod-Teller \cite{Axi1943} (AT) form 
\begin{equation}
u_{3}({\bf r}_{i},{\bf r}_{j},{\bf r}_{k})=\nu \frac{1+3\cos \theta _{i}\cos
\theta _{j}\cos \theta _{k}}{r_{ij}^{3}r_{ik}^{3}r_{jk}^{3}}  \label{EQAT}
\end{equation}
to take all these three-body contributions into account by modifying the
strength $\nu $ to include them empirically. In Eq. (\ref{EQAT}) $\theta
_{i} $, $\theta _{j}$ and $\theta _{k}$ denote, respectively, the angles at
vertex $i$, $j$ and $k$ of the triangle ($i,j,k$) with sides $r_{ij}=\left| 
{\bf r}_{j}-{\bf r}_{i}\right| $, $r_{ik}=\left| {\bf r}_{k}-{\bf r}%
_{i}\right| $ and $r_{jk}=\left| {\bf r}_{k}-{\bf r}_{j}\right| $. Usually,
the AT potential is designed to represent the $ddd$ contribution only with a
value $\nu \equiv \nu _{ddd}=2.204\times 10^{-26}$ J nm$^{9}$ as prescribed
by Leonard and Barker \cite{Leo1975}. Such an effective three-body potential
was recently used to study the liquid-vapor phase equilibria of argon \cite
{Miy1994} as well as the small-$k$ part of the static structure factor of
krypton in the dense liquid \cite{Gua2001}. For instance, to interpret
correctly their experimental data, Guarini {\it et al.} \cite{Gua2001} have
increased the strength $\nu $ by 65 \%, indicating that other three-body
contributions play an important role. Conversely, in a recent work \cite
{Bom2002}, we have shown that the missing contributions would have the
effect of slightly reducing the strength of the AT potential. Nevertheless,
these studies are not completely conclusive since calculations where carried
out within the integral equations theory, in which the three-body
contribution has to be treated as an approximate state-dependent effective
pair potential \cite{Cas1970}, which is not unique.

The purpose of this article is to analyze the influence of the three-body
interactions taken in the form of Eq. (\ref{EQAT}), to decide what is the
effective value of the strength $\nu $ for an accurate description of liquid
Kr. Three different situations are considered by varying $\nu $: a value of $%
\nu =0$ that corresponds to the pure two-body potential given by Aziz and
Slaman \cite{Azi1986}, a value of $\nu =\nu _{ddd}$ representing only the $%
ddd$ contribution, and a value $\nu _{eff}=1.65\nu _{ddd}$ which is believed
to reproduce the critical parameters of Kr within 1 \% \cite{Gua2001}.
Changing the strength of the AT potential is a convenient means to measure
of the departure from the $ddd$ term and, by comparison with the
experiments, could provide useful information on the contribution coming
from all other three-body terms that are omitted in the modelling of the
interaction between Kr atoms. This work represents an extension to the
liquid phase of our preceding studies on Kr in the gaseous phase \cite
{Bom1998,Jak2000b}. Thus, for a given interaction model, we carry out
molecular dynamics (MD) simulations to determine $S(k)$ as well as the
internal energy and the virial pressure. In this context, MD is a powerful
method \cite{All1989,Hai1992} since the three-body interactions can be
tackled without incurring the shortcomings of the approximate integral
equations. As a matter of fact, the treatment of the three-body potential (%
\ref{EQAT}) is not subject to arbitrariness since the resulting forces are
derived in an exact manner \cite{Hoh1981}, as for the two-body ones.

In order to extract a meaningful structure factor $S(k)$ from the MD
simulations, the pair correlation $g(r)$ is the key quantity that has to be
calculated as precisely as possible. Therefore, we have performed
large-scale MD in the sense that (i) a large enough simulation cell has been
considered so that $g(r)$ has a sufficient spatial extension to yield a
correct $S(k)$ by Fourier transform, especially at low $k$, and (ii) a
large\ number of time steps are produced in order to get a significant part
of the phase space trajectory, essential for the statistics. As three-body
forces are involved, this represents a huge amount of computer time, and we
have used a parallel algorithm described in some details in a previous work 
\cite{Jak2000}. Simulations with the different values of $\nu $ in the AT
potential have been performed in the microcanonical ($NVE$) ensemble with a
time step $\Delta t=5.67\cdot 10^{-15}$ s using $N=6912$ particles in a
cubic cell subject to the standard boundary conditions. The cutoff radius of
the interactions is $r_{c}=2.5r_{m}$, where $r_{m}=0.4008$ nm is the minimum
of the AS potential. For the three-body potential, this implies that
triplets of particles in which two or three distances of separations are
greater than $r_{c}$ are ignored in the calculation of the forces. In order
to investigate the influence of $r_{c}$ for the two-body potential, a value
of $4r_{m}$ is taken in specific cases. The typical duration of the runs is $%
113$ ps from which $1300$ independent configurations are extracted for the
statistical analysis of the physical quantities. Six states of liquid Kr
from the vicinity of the critical point to that of the triple point have
been studied, which are those investigated by Guarini {\it et al.} \cite
{Gua2001} and by Barocchi {\it et al.} \cite{Bar1993}, respectively at low-
and large-$k$. These states correspond to temperature $T=199$ K and
densities $n=12.10$, $11.66$ and $11.31$ nm$^{-3}$, $T=169$ K with $n=14.57$
and $14.22$ nm$^{-3}$, and $T=130$ K \ with $n=16.83$ nm$^{-3}$.

In Fig. 1, we present the large $k$ behavior of $S(k)$ for the three
different temperatures and we compare the MD curves, calculated with the
three different values of $\nu $, \ to the experimental data of Barocchi 
{\it et al.} \cite{Bar1993}. As the temperature increases and density
decreases, the first sharp diffraction peak as well as the subsequent
oscillations become less pronounced. This is well predicted by the MD
results since a remarkable agreement with the experiments is found whatever
the values of $\nu $. It can be seen that the three-body contributions have
only a minor influence even if a more careful examination shows that, at low
temperature, the height of the first peak is better described without the AT
potential while, at high temperature, a closer agreement is obtained when it
is included. At low-$k$, the three potential energy functions give rise to
completely different behaviors of $S(k)$ and the best results seem to be
those obtained with the $ddd$ interaction.

Let us now focus on the low-$k$ part of $S(k)$ in more details. We compare
the MD curves to the recent small-angle scattering data of Guarini {\it et
al.} \cite{Gua2001} along the $T=199$ K and $T=169$ K isotherms in Fig. 2
(a) and (b), respectively. At the highest temperature, near the critical
isotherm, a good agreement is found with the experiment at the three
densities with the $ddd$ potential. Examining the influence of $\nu $ at
density $n=12.10$ nm$^{-3}$, it appears that the AS two-body potential alone
($\nu =0$) is not able to predict the small-$k$ part of $S(k)$
satisfactorily, while the effective AT contribution ($\nu =\nu _{eff}$)
gives rise to a structure factor which is underestimated. The chain curve
corresponds to $S(k)$ obtained with the AS potential alone and a cutoff
radius $r_{c}=4r_{m}$. The influence of $r_{c}$ is seen only below $2.5$ nm$%
^{-1}$ and the curvature is slightly changed. As a result, this would have
the effect of increasing the values of $S(k)$ in this region whatever the
potential used. Taking $r_{c}=4r_{m}$ when the three-body contribution is
considered is computationally too costly and could represent a challenge for
future developments of the present MD code. However, for the three-body
potential, it is worth noting that, even with $r_{c}=2.5r_{m}$,
configurations in which one pair of particles of a triplet is separated up
to $5r_{m}$ are taken into account in the calculation of the forces. In
addition, the AT potential given by Eq. (\ref{EQAT}) decays as $r^{-9}$
therefore taking a cutoff radius larger than $2.5r_{m}$ seems not to be
necessary. The same observation can be drawn along the $T=169$ K isotherm,
as shown in Fig 2 (b), and in this case a very good agreement with the
experimental curves is seen when the $ddd$ potential is taken into account.
Interesting enough, both experimental data sets of Guarini {\it et al}. \cite
{Gua2001} and Barocchi {\it et al.} \cite{Bar1993}, which connect to each
other very well around $4$ nm$^{-1}$, lie between the curves corresponding
to $\nu =\nu _{ddd}$ and $\nu =\nu _{eff}$, for $k$ values in the range
between $2.5$ and $6.5$ nm$^{-1}.$

At $T=130$ K the situation is less clear as it can be seen in Fig. 3. While
the AS potential is also not sufficient to predict the small-$k$ behavior of 
$S(k)$, this time the best concordance with the experiment is obtained by
using $\nu =\nu _{eff}$. Nevertheless, the latter effective AT contribution
would have a tendency to underestimate the $PVT$ data while the $ddd$
strength gives the best prediction. In this case, taking again a cutoff
radius of $4r_{m}$ for the two-body potential alone (chain curve) has only a
negligible influence on $S(k)$, therefore our results for the two- plus
three-body might be correct. It should be stressed that the amplitude of $%
S(k)$ is very small at such a low temperature and high density state, and
the relative difference between the MD curves with $\nu _{ddd}$ and $\nu
_{eff}$ is of the same order of magnitude than the dispersion of the
experimental data points, which is about 10 \%. At this stage, we refrain
from drawing any conclusion and it would be desirable to dispose of an
accurate small angle scattering experiment for this thermodynamic state.

We also examine the influence of the three-body potential on the internal
energy $E$ and the virial pressure $P$ gathered in Table 1. Long-range
corrections have been applied, due to the truncation of the AS and the AT
potentials during the simulation at $r_{c}=2.5r_{m}$. The corrections to the
two-body part of the energy and the pressure are respectively $%
-0.153nr_{m}^{3}$ and $-0.312\left( nr_{m}^{3}\right) ^{2}$. For the
three-body parts, the corrections are estimated numerically using the
integral equation method \cite{Jak2000b}. For the energy, we obtain $%
0.007\left( nr_{m}^{3}\right) ^{2}$ for $\nu _{ddd}$ and $0.012\left(
nr_{m}^{3}\right) ^{2}$ for $\nu _{eff}$, while for the pressure we get $%
0.031\left( nr_{m}^{3}\right) ^{3}$ for $\nu _{ddd}$ and $0.051\left(
nr_{m}^{3}\right) ^{3}$ for $\nu _{eff}$. These quantities are expressed in
the unit of the minimum of the AS potential $\varepsilon =2.777\cdot
10^{-21} $ J. The influence of $\nu $ on the energy is moderate and does not
exceed $9 $ \% with respect to that of the two-body potential, even with $%
\nu _{eff}$. On the contrary, it has a large effect on the pressure. For the
two-body potential alone, the pressure is always negative, while by
including the three-body potential contribution, it becomes positive in the
majority of cases. In addition, $\nu _{eff}$ yields pressure values which
are too high, while the $ddd$ strength gives the best predictions, even if
it is always smaller compared to the experimental data. Again, the
calculated pressures with $\nu _{ddd}$ and $\nu _{eff}$ enclose the
experimental values.

Regarding the results presented above, it appears that the details of the
interaction model has no significant influence on $S(k)$ at large $k$, as
shown in Fig. 1. Even the AS two-body potential alone is able to reproduce
the structure factor with a good degree of accuracy. For the structure
factor at small scattering angle and the pressure, the three-body
interactions cannot be ignored in the liquid state. Moreover, it is seen
that the triple-dipole contribution gives the best agreement with the
experiments and therefore represents the main three-body effect. A more
precise examination shows that the structural and thermodynamic properties
depart substantially from the experiments even with the model combining the
AS two-body potential plus $ddd$ contribution. Indeed, discrepancies remain
(i) on the small $k$ part of $S(k)$ in the range between $2.5$ and $6.5$ nm$%
^{-1}$, where the calculated values are higher that the experimental ones,
especially for isotherms $T=169$ K and $T=130$ K, and (ii) on the pressure
where the theoretical results are systematically below the measurements, and
even remain negative for two thermodynamic states. Since the pressure varies
linearly with $\nu $, as it can be seen in Table 1, better results should be
obtained by increasing the value of $\nu _{eff}$ between $1.20\nu _{ddd}$
and $1.25\nu _{ddd}$, whatever the temperature.

We are led to the same conclusion as Guarini {\it et al. }\cite{Gua2001}{\it %
\ }that the value of $\nu $ in Eq. (\ref{EQAT}) has to be increased with
respect to that of the triple-dipole $\nu _{ddd}$. However, by the light of
the present MD calculations, an effective strength $\nu _{eff}=1.65\nu
_{ddd} $ seems to be too important, and we estimate that the additional
three-body terms beyond the $ddd$ one should represent up to $25$ \% of it.
Now the question arises to know what is the nature of the missing
contributions that will take a non negligible part in the interaction model.
According to Copeland and Kestner \cite{Cop1968} who studied liquid argon,
two majors three-body potentials beyond the $ddd$ one play an important
role, namely the exchange overlap and the $ddq$ dispersion acting
respectively at short and long interatomic distances. Both potentials are
known to have significant influence \cite{Lou1988,Bom2001}, however, while
the $ddq$ potential has the same sign as the $ddd$ contribution, the
exchange one has an opposite sign. Therefore, it will be of primary
importance to investigate their interplay in liquid krypton and whether they
improve the description of the interactions in liquid Kr when added to the $%
ddd$ term. It has been recently shown by van der Hoef and Madden \cite
{Hoef1999} in their simulation study of liquid argon that the $ddq$
contribution on the pressure is small but not negligible and it would be
interesting to extend these results in the case of Kr, not only for the
pressure but also for the structure factor.

At very small wave-number, i.e. $k<2.5$ nm$^{-1}$, the MD results of $S(k)$
underestimate the experimental curves as well as the $PVT$ data, which is
particularly visible along the isotherm $T=199$ K displayed in Fig. 2 (a).
As MD simulations are concerned, we are unavoidably confronted to finite
size effects and this fact might be attributed mainly to two factors: the
truncation of the pair-correlation functions at the half of the box size,
and the use of a cutoff radius of the interactions. Therefore, the $S(k)$
calculated by Fourier transform are subject to large uncertainties and must
be taken cautiously. Moreover, near the critical region, the correlation
length can exceed the size of the simulation box and for this reason the
present MD simulations might be not able to catch the correct behavior of $%
S(k)$ at $T=199$ K. As a matter of fact, as pointed out by Wilding \cite
{Wil1997} and Rovere \cite{Rov1993}, in the thermodynamic limit, critical
phenomena, like the divergence of $S(0)$, are smeared out and shifted.

In conclusion, MD simulations have been carried out for Kr in the liquid
phase for which new small angle scattering experiments were recently
performed \cite{Gua2001,Bar1993}. This study completes preceding works on
the low density and high temperature states of Kr \cite
{Bom1998,Jak2000b,Jak2000}, where it was demonstrated that the Aziz and
Slaman two-body potential associated to the Axirod and Teller triple-dipole
contribution gives a excellent representation of interactions in the gaseous
phase. Here, we have shown that the latter potential energy function
predicts the essential characteristics of structural properties in liquid
Kr, even though we believe that an accurate description of $S(k)$ at low $k$
and the thermodynamic properties requires that other three-body
contributions such as the dipole-dipole-quadrupole and the exchange overlap
potentials are taken into account. Large scale MD simulations including
these additional contributions will be performed in the near future.

The CINES (Centre Informatique National de l'Enseignement Sup\'{e}rieur) is
gratefully acknowledged for providing us with computer time under Project No
TMC1928. The authors would also like to thank Pr. R. Magli and Dr. E.
Guarini for providing us with the experimental data.

\newpage

\section*{Captions}

\begin{description}
\item  {\bf Figure 1. }Structure factor $S(q)$ for $T=130$ K, at $n=16.83$ nm%
$^{-3}$, for $T=169$ K, at $n=14.57$ nm$^{-3}$ and for $T=199$ K, at $%
n=12.10 $ nm$^{-3}$ from the top to the bottom (the curves for $T=169$ K and 
$T=130$ K are shifted upwards by an amount of $1$ and $2$, respectively),
calculated by molecular dynamics with $\nu =0$ (dashed lines) $\nu =\nu
_{ddd}$ (solid lines) and $\nu =\nu _{eff}$ (dotted lines) as described in
the text. Open circles correspond to the experimental data of Ref. \ref
{Bar1993}, while full circles stand for the $PVT$ data of Ref. \ref{Juz1976}.

\item  {\bf Figure 2.} Structure factor $S(q)$ at small scattering angle for
isotherm (a) $T=199$ K, at $n=12.10$ nm$^{-3}$, $n=11.66$ nm$^{-3}$ and $%
n=11.31$ nm$^{-3}$, and (b) $T=169$ K, at $n=14.57$ nm$^{-3}$.and $14.22$ nm$%
^{-3}$. The curves for $n=11.66$ nm$^{-3}$ and $n=11.31$ nm$^{-3}$ are
shifted upwards by an amount of $0.5$ and $1$, respectively, and that of $%
14.22$ nm$^{-3}$ by an amount of $0.2$. Molecular dynamics results are
carried out with $\nu =0$ (dashed lines), $\nu =\nu _{ddd}$ (solid lines)
and $\nu =\nu _{eff}$ (dotted lines) as described in the text. The chain
curve corresponds to MD results with $\nu =0$ and a cutoff radius of $4r_{m}$%
. Open circles correspond to the experimental data of Ref. \ref{Bar1993},
open triangles correspond to the experimental data of Ref. \ref{Gua2001} and
full circles stand for the $PVT$ data of Ref. \ref{Juz1976}.

\item  {\bf Figure 3. }Structure factor $S(q)$ at small scattering angle for 
$T=130$ K, at $n=16.83$ nm$^{-3}$. Same captions as in Fig.1.

\item  {\bf Table 1}. Excess internal energy, $E^{\text{ex}}/N\varepsilon $,
and pressure, $P\left( r_{m}\right) ^{3}/\varepsilon $, calculated by
molecular dynamics for the different thermodynamic states considered in this
work. Subscripts 2 and 3 stand respectively for the two- and two- plus
three-body parts of the internal energy. $P_{\text{exp}}$ corresponds to the
experimental values of Ref. \ref{Gua2001}. The numbers in brackets represent
the standard deviations that affect the last decimal of the temperature,
energy and pressure extracted from the simulation.
\end{description}

\newpage

\section*{Table}

\begin{center}
\hspace*{-0.5in}\bigskip 
\begin{tabular}{rrrrrrrr}
\hline\hline
$T$ (K) & $\rho $ (nm$^{-3}$) & $\nu /\nu _{ddd}$ & $E_{2}^{\text{ex}%
}/N\varepsilon $ & $E_{3}^{\text{ex}}/N\varepsilon $ & $Pr_{m}^{3}/%
\varepsilon $ & $P$ (bar) & $P_{\text{exp}}$ (bar) \\ \hline
130{\small \ (1)} & 16.83 & $0$ & $-4.962${\small \ (6)} & 
\multicolumn{1}{c}{$-$} & $-0.87${\small \ (4)} & $-375.4$ &  \\ 
&  & 1 & $-4.953${\small \ (6)} & $-4.689${\small \ (6)} & $-0.12${\small \
(4)} & $-51.8$ & 0 \\ 
&  & 1.65 & $-4.958${\small \ (6)} & $-4.522${\small \ (6)} & $0.33$ {\small %
(4)} & $142.4$ &  \\ \hline
169{\small \ (1)} & 14.57 & 0 & \multicolumn{1}{l}{$-4.151${\small \ (7)}} & 
\multicolumn{1}{c}{$-$} & \multicolumn{1}{l}{$-0.37${\small \ (5)}} & $%
-159.6 $ &  \\ 
&  & 1 & $-4.127${\small \ (7)} & $-3.938${\small \ (7)} & $0.06${\small \
(4)} & $25.9$ & $61.7$ \\ 
&  & 1.65 & $-4.126${\small \ (7)} & $-3.818${\small \ (7)} & $0.30${\small %
\ (4)} & $129.4$ &  \\ \hline
169{\small \ (1)} & 14.22 & 1 & \multicolumn{1}{l}{$-4.034${\small \ (7)}} & 
\multicolumn{1}{l}{$-3.855${\small \ (7)}} & $-0.05$ {\small (4)} & $-21.5$
& $20.1$ \\ \hline
199{\small \ (1)} & 12.10 & 0 & $-3.392${\small \ (8)} & \multicolumn{1}{c}{$%
-$} & $-0.06${\small \ (4)} & $-25.9$ &  \\ 
&  & 1 & $-3.358${\small \ (9)} & $-3.232${\small \ (9)} & $0.11${\small \
(4)} & $49.6$ & $73.1$ \\ 
&  & 1.65 & $-3.344${\small \ (8)} & $-3.138${\small \ (8)} & $0.27${\small %
\ (4)} & $116.5$ &  \\ \hline
199{\small \ (1)} & 11.66 & 1 & $-3.245${\small \ (8)} & $-3.127${\small \
(8)} & $0.09${\small \ (4)} & $38.8$ & $55.7$ \\ \hline
199{\small \ (1)} & 11.31 & 1 & \multicolumn{1}{l}{$-3.165${\small \ (8)}} & 
\multicolumn{1}{l}{$-3.053${\small \ (8)}} & $0.06${\small \ (4)} & $25.9$ & 
$46.3$ \\ \hline\hline
\end{tabular}

{\bf Table 1}
\end{center}


\newpage

\begin{references}
\bibitem{Bar1976}  J. A. Barker and D. Henderson, Rev. Mod. Phys. {\bf 48},
589 (1976).

\bibitem{End1965}  J. E. Enderby, T Gaskell and H. H. March, Proc. Phys Soc.
London. {\bf 85}, 217 (1965).

\bibitem{Rea1992}  L. Reatto and M. Tau, J. Phys.: Condens. Matter {\bf 4},
1 (1992).

\bibitem{Mag1996}  R. Magli, F. Barocchi, P. Chieux, R. Fontana, Phys. Rev.
Lett. {\bf 77}, 846 (1996).

\bibitem{For1997}  F. Formisano, C. J. Benmore, U. Bafile, F. Barocchi, P.
A. Egelstaff, R. Magli and P. Verkerk, Phys. Rev. Lett. {\bf 79}, 221 (1997).

\bibitem{For1998}  F. Formisano, F. Barocchi and R. Magli, Phys. Rev. E {\bf %
58}, 2648 (1998).

\bibitem{Ben1999}  \label{Ben1999}C. J. Benmore, F. Formisano, R. Magli, U.
Bafile, P. Verkerk, P. A. Egelstaff, and F. Barocchi, J. Phys.: Condens.
Matter {\bf 11}, 3091(1999).

\bibitem{Gua2001}  \label{Gua2001}E. Guarini, R. Magli, M. Tau, F. Barocchi,
G. Casanova and L. Reatto, Phys. Rev. E {\bf 63}, 052201 (2001).

\bibitem{Lon1930}  F. London, Z. Phys. {\bf 63}, 245 (1930).

\bibitem{Pow2001}  E. A. Power, Eur. J. Phys., {\bf 22}, 453 (2001).

\bibitem{Cas1948}  H. B. G. Casimir and D. Polder, Phys. Rev. {\bf 73}, 360
(1948).

\bibitem{Azi1985}  R. A. Aziz and M. J. Slaman, Mol. Phys. {\bf 57}, 827
(1985).

\bibitem{Azi1986}  R. A. Aziz and M. J. Slaman, Mol. Phys. {\bf 58}, 679
(1986).

\bibitem{Axi1943}  B. M. Axilrod and E. Teller, J. Chem. Phys. {\bf 11}, 299
(1943).

\bibitem{Bel1970}  R. J. Bell, J. Phys. B {\bf 3}, 731 (1970).

\bibitem{Cop1968}  D. A. Copeland and N. R. Kestner, J. Chem. Phys. {\bf 49}%
, 5214 (1968).

\bibitem{Lou1988}  P. Loubeyre, Phys. Rev. B {\bf 37}, 5432 (1988).

\bibitem{Sad1996}  R. J. Sadus and J. M. Prausnitz, J. Chem. Phys. {\bf 104}%
, 4784 (1996).

\bibitem{Lot1997}  V. F. Lotrich and K. Szalewicz, Phys. Rev. Lett. {\bf 79}%
, 1301 (1997).

\bibitem{Leo1975}  P. J. Leonard and J. A. Barker, Theor. Chem. Adv.
Perspect. {\bf 1}, 117 (1975).

\bibitem{Miy1994}  Y. Miyano, Fluid Phase Equilib. {\bf 95}, 31 (1994).

\bibitem{Bom2002}  J. M. Bomont, N. Jakse and J.\ L. Bretonnet (to be
published).

\bibitem{Cas1970}  G. Casanova, R. J. Dulla, D. A. Johan, J. S. Rowlinson
and G. Savile, Mol. Phys. {\bf 18}, 589 (1970).

\bibitem{All1989}  M. P. Allen and D. J. Tildesley, ''{\em Computer
Simulation of liquids}`` (Clarendon Press 1989).

\bibitem{Hai1992}  J. M. Haile, ''{\em Molecular Dynamics Simulation :
Elementary Methods``}, ed. John Wiley and Sons Inc. (1992).

\bibitem{Hoh1981}  C. Hoheisel, Phys. Rev.{\em \ }A {\bf 23,} 1998 (1981).

\bibitem{Bom1998}  J. M. Bomont, N. Jakse and J. L. Bretonnet, Phys. Rev.%
{\bf \ }B{\bf \ 57}, 10217 (1998).

\bibitem{Jak2000b}  N. Jakse, J. M. Bomont, I. Charpentier and J. L.
Bretonnet, Phys. Rev.{\bf \ }E{\bf \ 62}, 3671 (2000).

\bibitem{Jak2000}  N. Jakse and I. Charpentier, Mol. Sim. {\bf 23}, 293
(2000).

\bibitem{Bar1993}  \label{Bar1993}F. Barocchi, P. Chieux, R Magli, L. Reatto
and M. Tau, J. Phys.: Condens. Matter {\bf 5}, 42991(1993).

\bibitem{Bom2001}  J. M. Bomont, J. L. Bretonnet and M. A. van der Hoef, J.
Chem. Phys. {\bf 114}, 5674 (2001).

\bibitem{Hoef1999}  M. A. van der Hoef and P. A. Madden, J. Chem. Phys. {\bf %
111}, 1520 (1999).

\bibitem{Juz1976}  \label{Juz1976}J. Juza and O. Sifner, Acta Technica CSAV 
{\bf 1}, 1 (1976).

\bibitem{Wil1997}  N. B. Wilding, J. Phys.: Condens. Matter {\bf 9}, 585
(1997).

\bibitem{Rov1993}  M. Rovere, J. Phys.: Condens. Matter {\bf 5}, B193 (1993).
\end{references}
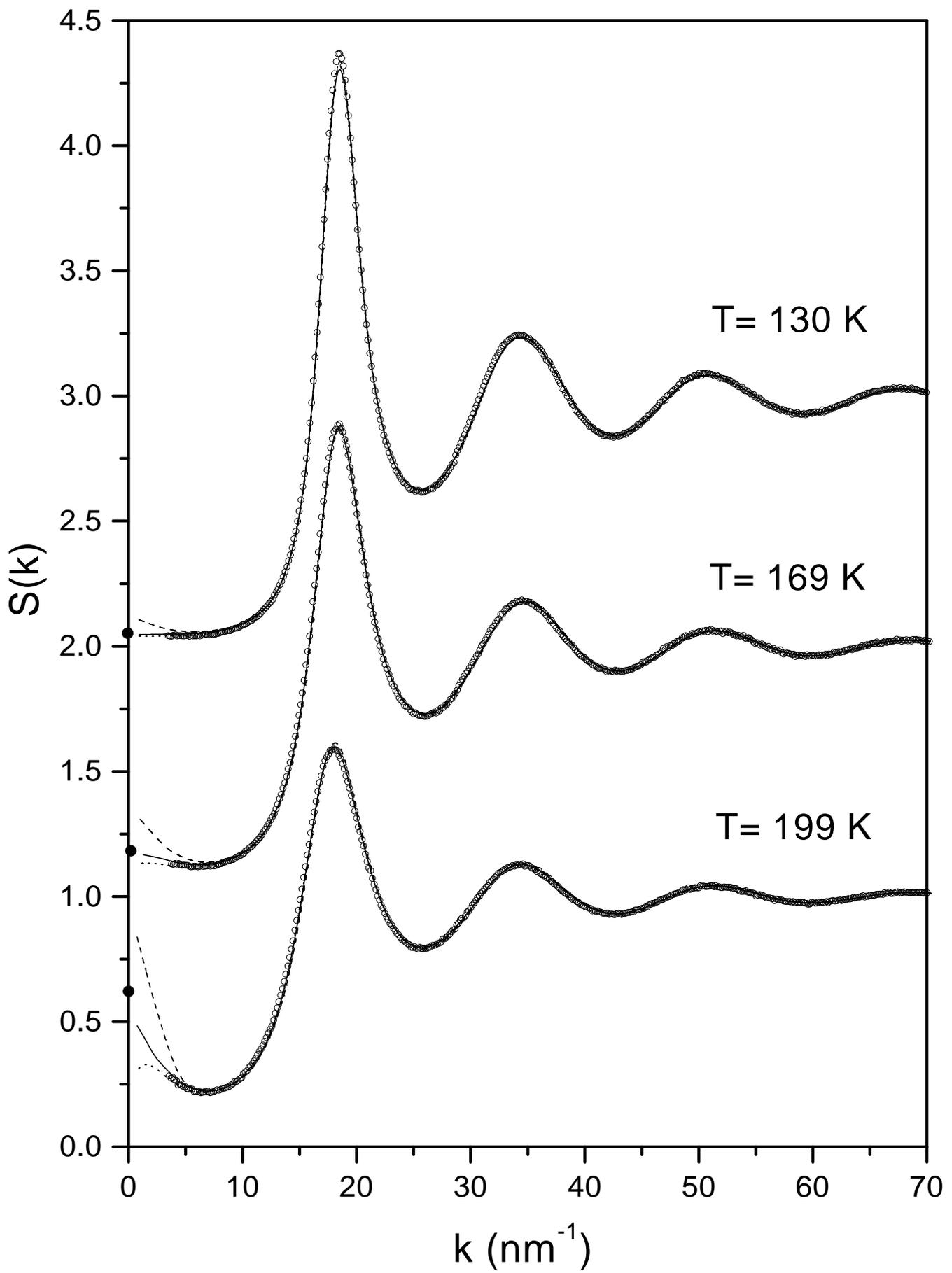
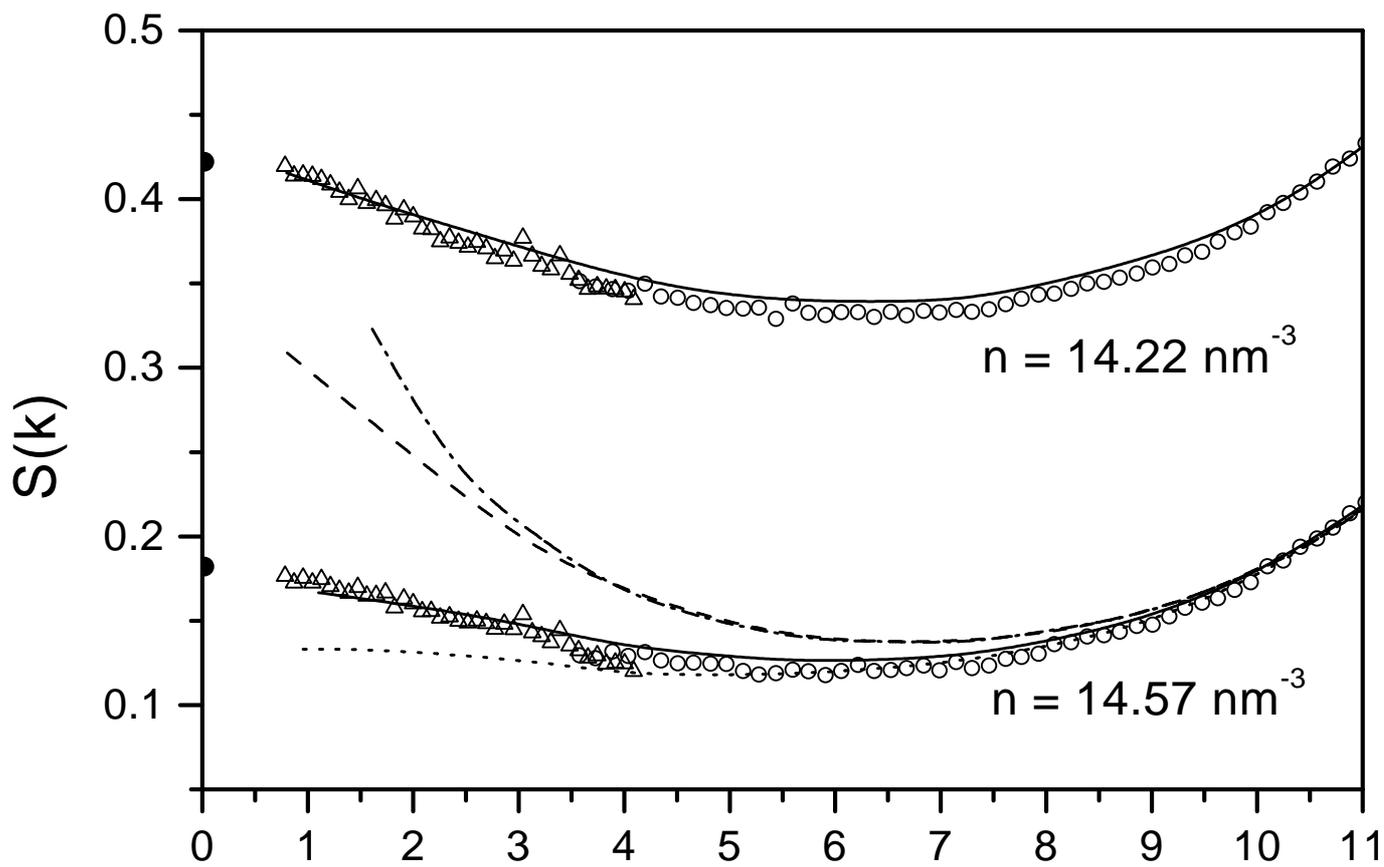
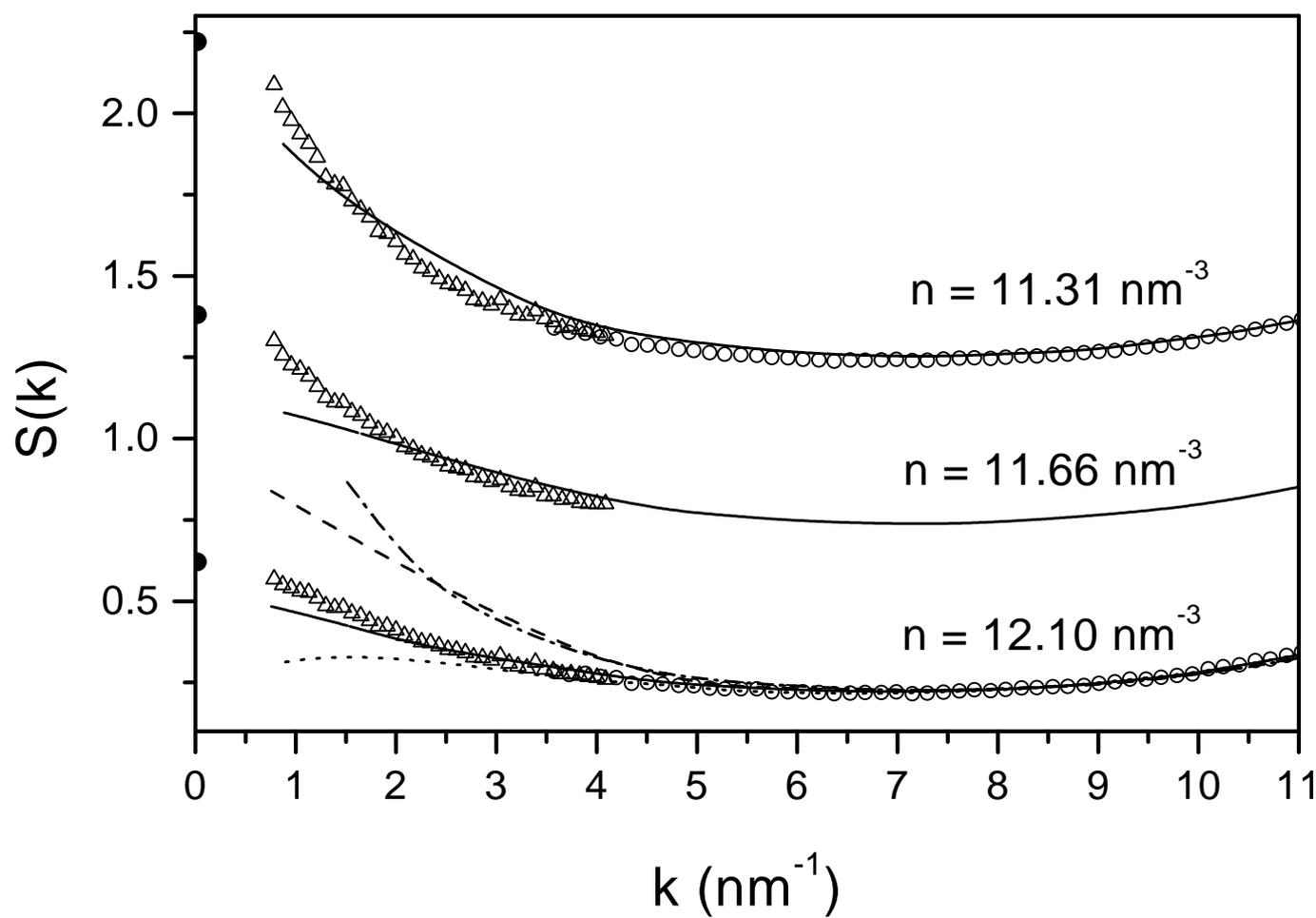
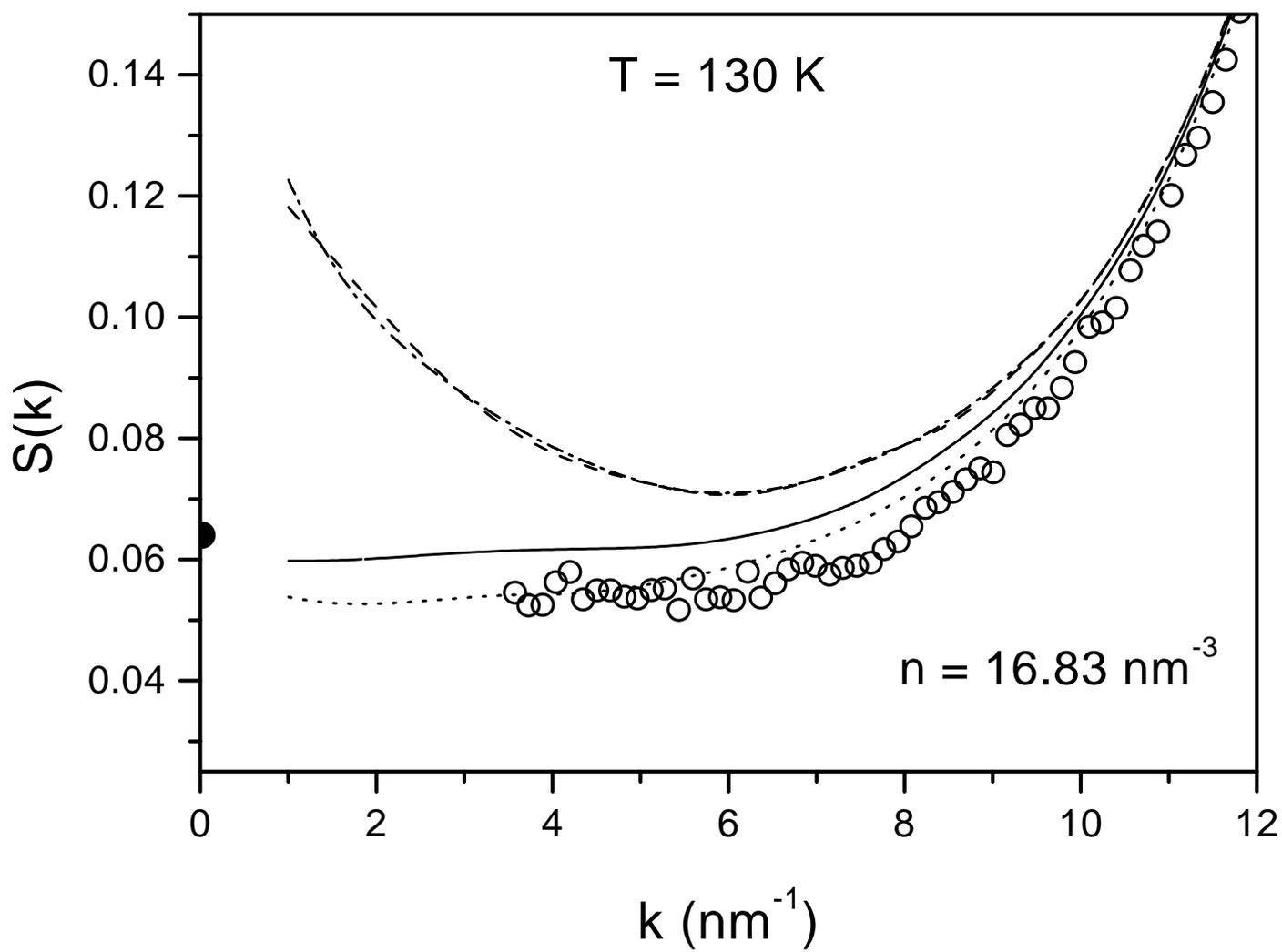
\end{document}